\documentstyle[epsfig,floats,twocolumn,eqsecnum,aps]{revtex}

\begin{document}
\draft \preprint{HEP/123-qed}
\renewcommand{\textfraction}{0.1}
%\renewcommand{\topfraction}{0.9}
%\psfigurepath={./:./.}

\wideabs{

\title {Superconductivity on the threshold of magnetism in
        CePd$_2$Si$_2$ and CeIn$_3$} \author
        {F. M. Grosche\cite{byline}, I. R. Walker, S. R. Julian,
        N. D. Mathur, D. M. Freye, M. J. Steiner and G. G. Lonzarich}

\address {Cavendish Laboratory, Madingley Road, Cambridge CB3
        0HE, UK} \date{\today} \maketitle

\begin{abstract}
The magnetic ordering temperature of some rare earth based heavy
fermion compounds is strongly pressure-dependent and can be completely
suppressed at a critical pressure, p$_c$, making way for novel
correlated electron states close to this quantum critical point. We
have studied the clean heavy fermion antiferromagnets CePd$_2$Si$_2$
and CeIn$_3$ in a series of resistivity measurements at high pressures
up to 3.2 GPa and down to temperatures in the mK region. In both
materials, superconductivity appears in a small window of a few tenths
of a GPa on either side of p$_c$. We present detailed measurements of
the superconducting and magnetic temperature-pressure phase diagram,
which indicate that superconductivity in these materials is enhanced,
rather than suppressed, by the closeness to magnetic order.
\end{abstract}

}

\narrowtext

\section {Introduction}
\label{sec:level1}
The intense Coulomb interaction and low bandwidth driving electronic
correlations in heavy fermion systems lead to a fascinating variety of
low temperature states that remain only partly understood.
%Reminiscent of $^3$He, but at a new level of complexity and diversity,
%these compounds hold the potential for many types of spin-aligned and
%superfluid order, as well as unconventional normal states.  
In analogy with liquid $^3$He, but at a new level of complexity and
diversity, the effective quasiparticle interaction responsible for the
high scattering rates and effective masses of these systems could for
suitable quasiparticle spin-orientation become attractive enough to
foster unconventional superconductivity.

Increased and widespread interest in these compounds arose, when
superconductivity was indeed discovered in the heavy fermion compound
CeCu$_2$Si$_2$ \cite{steglich79}, which is now known to rest on the
border of magnetic order at ambient pressure, and can be moved into
the magnetic state by suitable sample preparation or doping
\cite{trovarelli97,knebel96,gegenwart98}. Further convincing evidence
for the closeness of this material to magnetic order emerges from high
pressure measurements on the isoelectronic and isostructural relative
CeCu$_2$Ge$_2$ \cite{jaccard92}. Magnetism makes way in this material
to superconductivity at a pressure just sufficient to reduce the unit
cell volume to that of CeCu$_2$Si$_2$ at ambient pressure.

The close association of superconductivity with the vanishing of a
magnetic phase in CeCu$_2$Si$_2$/Ge$_2$ prompts the question whether
the two types of order are simply in competition, as would be
suggested for a conventional pairing scenario, or whether they could
be related, as is suggested by the analogy with liquid $^3$He. In the
former case, the proximity of the two phases would be purely
accidental, whereas in the latter an attractive magnetic component of
the quasi-particle interaction could contribute to forming a
superconducting state close to the threshold of magnetism. In this
{\em magnetic interaction} scenario, spin fluctuations effectively
replace lattice vibrations in binding the heavy quasiparticles into
Cooper pairs. As this interaction is dominant only close to the border
of magnetism, one might expect to find superconductivity typically
only over a narrow region in lattice density, connected to the
disappearance of magnetic order and extending to both sides of the
critical point, at least when the transition is continuous.  By
contrast, the superconducting region in CeCu$_2$Si$_2$/Ge$_2$ extends
over nearly 10 GPa in pressure to the high-pressure side of the
critical point, suggesting a more complicated origin for
superconductivity in this case.

This observation and the extensively studied metallurgical
peculiarities of the CeCu$_2$Si$_2$ system motivate a search for
related but simpler materials.  Strikingly, nearly two decades of
research into Ce-based heavy fermion compounds failed to bring up any
other superconductors, raising the question, why superconductivity
should be so narrowly confined to the CeCu$_2$Si$_2$/Ge$_2$ system.

Within the magnetic interaction picture, one may expect to find such
candidates by tuning magnetically ordered compounds through the point
where the ordering temperature falls continuously to zero. The study
of these {\em quantum critical points} has a history dating back to
theoretical and experimental work on d-metals
\cite{lonzarich97,moriya85,hertz76,pfleiderer97,ishikawa85,lonzarich85,lonzarich86,grosche95,bernhoeft83,bernhoeft88,lonzarich89}.
Recently, the systematic investigation of the magnetic quantum
critical point in Ce-based systems has brought to light a host of pure
compounds, which show a tendency towards superconductivity when they
are tuned through the threshold of magnetism as a function of lattice
density by means of hydrostatic pressure.  (CePd$_2$Si$_2$
\cite{grosche96}, CeRh$_2$Si$_2$ \cite{movshovich96}, CeCu$_2$
\cite{vargoz97}, CeIn$_3$ \cite{walker97} and possibly also CeCu$_5$Au
\cite{wilhelm99}).

\begin{figure} [tb]
\centerline{\epsfig{figure=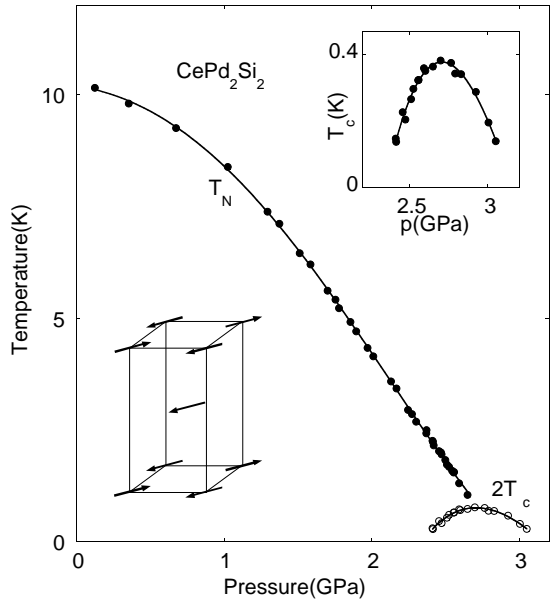,width=3.2in,clip=on}}
\par
\caption{Pressure-temperature phase diagram of
CePd$_2$Si$_2$. Magnetic ($T_N$) and superconducting ($T_c$)
transition temperatures have been determined from the mid-points of
$d\rho/dT$ and $\rho$, respectively. The inset shows the
superconducting part of the phase diagram in more detail. As can be
inferred from Fig. \protect\ref{cpssupress}, the transition width, $\Delta
T_c$, is typically of the order of 50 mK, but rises towards the high-
and low-pressure borders of the superconducting region.}
\label{cpsphases}
\end{figure}

\begin{figure}[tb]
\centerline{\epsfig{figure=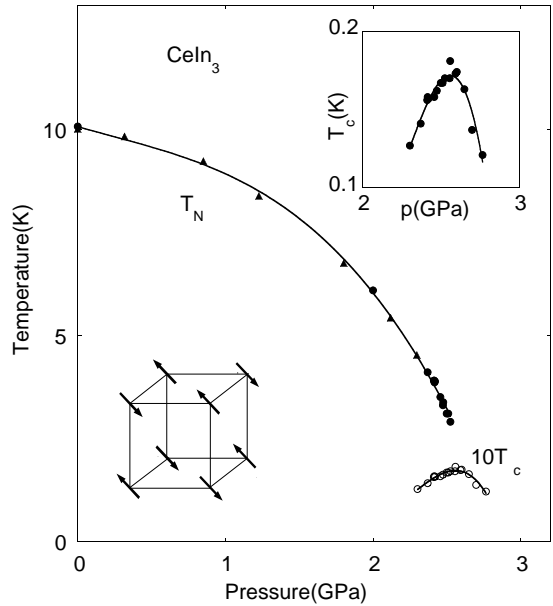,width=3.2in,clip=on}}
\par
\caption{Pressure-temperature phase diagram of CeIn$_3$. Magnetic
($T_N$) and superconducting ($T_c$) transition temperatures have been
determined from the mid-points of $d\rho/dT$ and $\rho$,
respectively. The solid triangles have been taken from \protect\cite{morin88},
scaling the pressure-axis by a fixed factor of about 90\%. The
different absolute pressure scale evident from \protect\cite{morin88} is
attributed to differences in the pressure calibration.  The inset
shows the superconducting part of the phase diagram in more detail.
The transition width, $\Delta T_c$, is only weakly pressure dependent
and remains smaller than 20 mK where it could be determined.}
\label{cinphases}
\end{figure}

In this paper, we concentrate on two of these materials, the
tetragonal CePd$_2$Si$_2$, which was the first Ce-based heavy fermion
superconductor to be found outside the CeCu$_2$Ge$_2$/Si$_2$ system,
and the cubic CeIn$_3$. In both materials we find a narrow
superconducting region of only a few tenths of a GPa around the
critical pressure, $p_c$, where the antiferromagnetic ordering
temperature extrapolates to zero. Detailed investigations of their
normal and superconducting properties, in particular by neutron
scattering, are complicated by the need to use hydrostatic pressure to
suppress magnetic order.  Fortunately, CeNi$_2$Ge$_2$, the
isoelectronic relative of CePd$_2$Si$_2$ has been found to display at
ambient pressure a behaviour analogous to that of CePd$_2$Si$_2$ just
beyond the critical pressure \cite{knebel99,grosche98,grosche00}. Studies on
CeNi$_2$Ge$_2$ enable us to extend the phase diagram of CePd$_2$Si$_2$
to higher effective pressure and to examine the evolution of the
normal state and novel ordered states in greater detail
\cite{grosche98,grosche00,grosche97,lister96,agarwal00,braithwaite00}.
The resulting phase diagram shows not only a low pressure
superconducting phase, limited to a small pressure range similar to
that observed in CePd$_2$Si$_2$, but also an unexplored magnetic phase
arising at high pressures of the order of 1.5 GPa and coinciding with
a further superconducting region at high pressure, which extends up to
at least about 3 GPa. This leads us to expect a similar second
superconducting region in the sister comound CePd$_2$Si$_2$ starting
at pressures about 1.5 GPa above $p_c$, i.e above 4 GPa and stretching
up to about 3 GPa above $p_c$, up to about 6 GPa.  Unfavourable sample
or pressure conditions may in some cases cause the two superconducting
ranges to merge, giving rise to the impression of a single, wide
superconducting phase.

The merging of two superconducting ranges under non-hydrostatic
pressure would offer an explanation for the drastic contrast between
Bridgman-cell measurements on CePd$_2$Si$_2$ using a solid pressure
medium \cite{raymond99}, which show a wide superconducting region
above $p_c$, and both our original determination of the phase diagram
in a hydrostatic pressure cell \cite{mathur98} and its recent
independent confirmation \cite{sheikin2000,demuer2000}. While unsuitable for
detailed studies of the critical region, where under {\em hydrostatic}
conditions, a very narrow superconducting region can clearly be
resolved, the Bridgman-cell measurements give a first indication of
the possible presence of a second high pressure superconducting phase
in CePd$_2$Si$_2$, as anticipated from earlier findings on
CeNi$_2$Ge$_2$
\cite{grosche98,grosche00,grosche97,lister96,agarwal00,braithwaite00}.

These findings on CeNi$_2$Ge$_2$ and CePd$_2$Si$_2$ raise the
intriguing possibility that some tetragonal Cerium heavy fermion
compounds could be nearly critical over a wide region in parameter
space, pressure or composition, maybe as a result of the competition
between different types of magnetic order. This scenario may present
the simplest and most intuitive explanation for the surprising
stability of superconductivity in CeCu$_2$Si$_2$.

\subsection {CePd$_2$Si$_2$ and CeIn$_3$}

The heavy fermion antiferromagnets CePd$_2$Si$_2$ and CeIn$_3$ offer
an interesting opportunity for studies of the antiferromagnetic heavy
fermion critical point.  CePd$_2$Si$_2$ is isostructural to the heavy
fermion superconductor CeCu$_2$Si$_2$ \cite{steglich79} and its larger
volume relative CeCu$_2$Ge$_2$ \cite{jaccard92} (ThCr$_2$Si$_2$
structure), but differs from CeCu$_2$Si$_2$ in the number of d
electrons in the d metal constituent, and hence in the character of
the Fermi surface and in the magnetic properties
\cite{movshovich96}. At ambient pressure, CePd$_2$Si$_2$ orders in an
antiferromagnetic structure below a N\'{e}el temperature \mbox{$T_N$}
of about 10 K with a comparatively small moment of $\simeq 0.7 \mu_B$
\cite {grier84}. The spin configuration consists of ferromagnetic
(110) planes with spins normal to the planes and alternating in
directions along the spin axis.  Previous high pressure measurements
indicate that T$_N$ is strongly pressure dependent in CePd$_2$Si$_2$
and has a critical pressure within the range of conventional
hydrostatic pressure cells.

The discovery of superconductivity in CeIn$_3$ provides a cubic
reference material, which may be simpler to describe theoretically.  
CeIn$_3$ crystallises in the cubic AuCu$_3$ structure
and orders at ambient pressure near 10 K with the ordering wavevector
in the (111) direction \cite{morin88}. 

Here, we describe the results of a detailed study of the critical
region, where the antiferromagnetic ordering temperature $T_N$ falls
to zero as a function of pressure, in both CePd$_2$Si$_2$ and CeIn$_3$
(Figs.~ \ref{cpsphases} and \ref{cinphases}). Both materials display a
very narrow superconducting region close to $p_c$, but the shapes of
the superconducting phase lines, $T_c (p)$ differ slightly. Moreover,
close to the critical pressure, the normal state resistivity, $\rho$,
in CeIn$_3$ approaches the low temperature limiting form $\Delta\rho
\equiv \rho - \rho_0 \propto T^{3/2}$, where $\rho_0$ denotes the
residual resistivity, smoothly. By contrast, in CePd$_2$Si$_2$ the
resistivity is nearly linear over almost two orders of magnitude in
temperature. Moving away from $p_c$, a Fermi-liquid $T^2$ resistivity
appears to return more rapidly in CeIn$_3$ than in CePd$_2$Si$_2$.

\section {Experimental Techniques}

We have used pressure as the exclusive control parameter in a study of
the stoichiometric heavy fermion compounds CePd$_2$Si$_2$ and CeIn$_3$
to minimise possible complications due to impurities or disorder. As
the critical pressure in these compounds lies within the range of
conventional piston-cylinder clamp cells, full advantage can be taken
of the high pressure homogeneity and hydrostaticity available with
this method, which makes possible a detailed study of the phase lines
$T_c(p)$ over a very narrow pressure range.  The electrical
resistivity at pressures up to about 3.2 GPa was measured by a low
power AC four-terminal method inside BeCu/Maraging Steel hydrostatic
clamp cells filled with a 1:1 mixture of iso-pentane and
n-pentane. The pressure was obtained to within $\pm 0.1 kbar$ from the
resistively determined superconducting transition temperature of a tin
sample \cite{smith69}. The sample resistance was scaled to published
values for the room temperature resistivity to obtain an absolute
resistivity scale \cite{resscales}.  Experiments were carried out on a
top-loading dilution refrigerator and an adiabatic demagnetisation
refrigerator.  Single crystals of CePd$_2$Si$_2$ and CeIn$_3$ were
prepared by a radio frequency melting technique in a water cooled
copper crucible and UHV chamber in which ultra-pure argon under
pressures of up to 8 bar was introduced for part of the
process. Further details of the preparation process and of the sample
characterisation are given in \cite{mathur95}. The residual
resistivity ratio $\rho(300K)/\rho(T \rightarrow 0K)$ is approximately
50 at atmospheric pressure for the best crystals of CePd$_2$Si$_2$ and
100 for CeIn$_3$.  Apart from the comparative study on the role of
sample quality shown in Fig.~\ref{cpsfieldandsamples}b, all the
results on CePd$_2$Si$_2$ were obtained on a single specimen with a
resistance ratio $RRR \equiv \rho(300 K)/\rho_0 \simeq 34$ ($\rho_0
\simeq 1.4 \mu\Omega cm$). A common problem in the preparation of
CeIn$_3$ is the inclusion of small amounts of Indium in the sample.
Care was taken to achieve vigorous stirring during the initial stages
of preparation. Samples were then checked for the presence of In
inclusions by resistivity measurements below 4K. The superconducting
phase diagram presented here was recorded on a sample with a minimal
amount of second phase, which is detected as an anomaly of order 1\%
in the resistivity trace.  A second sample with similar residual
resistivity ($0.6 \mu \Omega cm$ viz. $0.7 \mu \Omega cm$), but no
detectable Indium inclusion was used to study the normal state
properties, and to confirm the phase diagram at selected pressures.

%Extracting power-law exponents from the
%temperature-dependence of the resistivity has become a crucial tool
%for comparing the experimentally observed critical behaviour with
%theoretical calculations. Numerical methods for determining critical
%exponents are presented and evaluated in the appendix.

\section {Results}
% \label{sec:level1}

Resistivity measurements, while difficult to interpret quantitatively,
can give important qualitative information. In heavy fermion systems,
the high temperature resistivity is usually large ($\sim 50 \mu \Omega
cm$) and only weakly temperature dependent, consistent with the
scattering of charge carriers off thermally disordered moments. As the
temperature is lowered, this scattering begins to drop significantly
at an upper temperature scale, $T_{sf}$, below which thermal
fluctuations of the local magnetic order parameter are increasingly
frozen out. At low enough temperature, below a scale $T_{FL}$, most
materials approach the low temperature limiting forms predicted by
Fermi liquid theory, which should give an electronic contribution to
the resistivity $\Delta\rho \propto T^2$.  However, the spin degrees
of freedom may also condense into magnetically ordered states at
ordering temperatures $T_m$ ($T_N$ in the antiferromagnetic case),
leading to further decreases in magnetic scattering, and consequently
to anomalies or kinks in the resistivity trace at low temperatures. In
addition, the formation of an antiferromagnetically ordered state can
be accompanied by the opening of a charge gap over a section of the
Fermi surface, which may be visible as an increase in the resistivity
on cooling, when measured along particular crystal orientations.
Important information on the nature of the normal state can be gained
by examining the low temperature form of $\rho(T)$. In a number of
materials on the border to magnetism one finds $\Delta\rho \propto
T^x$ with $x < 2$. This is evidence for an anomalous quasiparticle
scattering mechanism, which appears to go beyond the regular
quasiparticle interaction on which conventional Fermi liquid theory is
based.

\subsection{CePd$_2$Si$_2$}

\begin{figure} [tb]
\centerline{\epsfig{figure=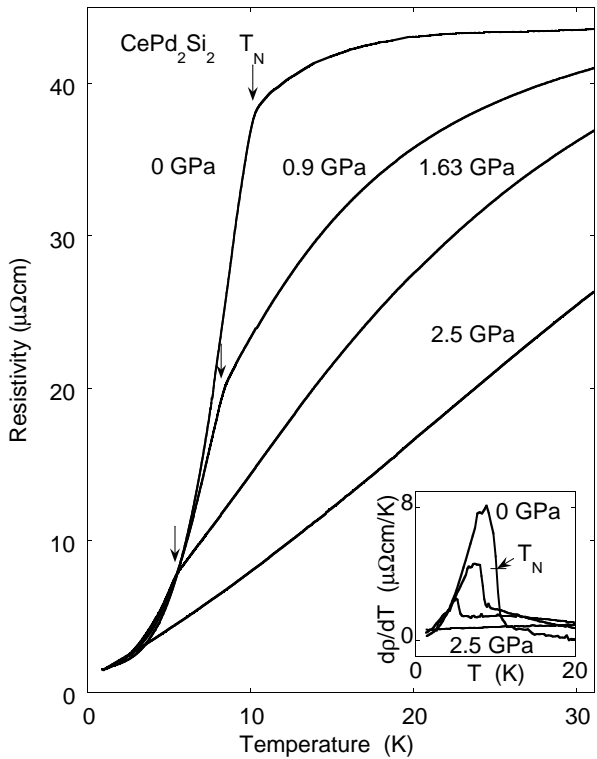,width=3.3in,clip=on}}
\par
\caption{The temperature dependence of the resistivity, $\rho$, of
CePd$_2$Si$_2$ measured along the a-axis of a sample with residual
resistance ratio $\mbox{RRR}=34$ at different pressures. The N\'eel
temperature, $T_N$, marked by arrows, is extracted as the
mid-point of the sudden change in the slope of $\rho(T)$,
$d\rho/dT$ (inset). }
\label{cpsresall}
\end{figure}

The resistivity of CePd$_2$Si$_2$ exhibits a very strong presssure
dependence (Fig.~\ref{cpsresall}). We can identify three main
features. Firstly, a sharp kink occurs in the resistivity at a
temperature T$_N$, which coincides, at zero pressure, with the
antiferromagnetic ordering temperature (or N\'{e}el temperature)
deduced from neutron experiments, $T_N (p=0) \simeq 10.2 K$
\cite{grier84}. The position of the kink in the resistivity is assumed
to correspond to the N\'{e}el temperature also at higher
pressures. T$_N$ decreases with increasing pressure and falls below 1
K at around 2.6 GPa.  Above about 1.5 GPa, this decrease is nearly
linear with a slope of -4.8 K/GPa, and extrapolates to a critical
pressure $p_c$ of 2.86 GPa (Fig.~\ref{cpsphases}).  This linear
pressure dependence $T_N \propto (p_c-p)$ is not consistent with
conventional spin-fluctuation predictions for a three-dimensional
antiferromagnet, where one would naively expect $T_N \propto
(p_c-p)^{z/d} = (p_c-p)^{2/3}$ \cite{lonzarich97}. Here, $z$ denotes
the dynamical exponent, usually assumed to be 2 in the
antiferromagnetic case, and $d$ is the dimensionality of the spin
fluctuation spectrum. A linear pressure dependence of $T_N$ would in
this model rather point towards $d = 2$, or at least indicate a very
anisotropic spin fluctuation spectrum.

%However, the wide range of this linear
%behaviour remains surprising in the simplest magnetic fluctuation
%approach.

Secondly, the behaviour of the resistivity above T$_N$ changes
dramatically with pressure.
At zero pressure, the resistivity stays constant to within 10\% from
room temperature down to the N\'{e}el temperature, below which it
falls sharply to a small residual value. At higher pressures, the
resistivity develops a progressively stronger temperature dependence
in the normal state. In particular, the temperature of the `shoulder',
below which the resistivity falls off, increases rapidly with
pressure. The resistivity curves for $T > T_N$ appear to scale roughly
with a characteristic temperature which we
associate with the spin fluctuation temperature $T_{sf}$.

\begin{figure} [bt]
\centerline{\epsfig{figure=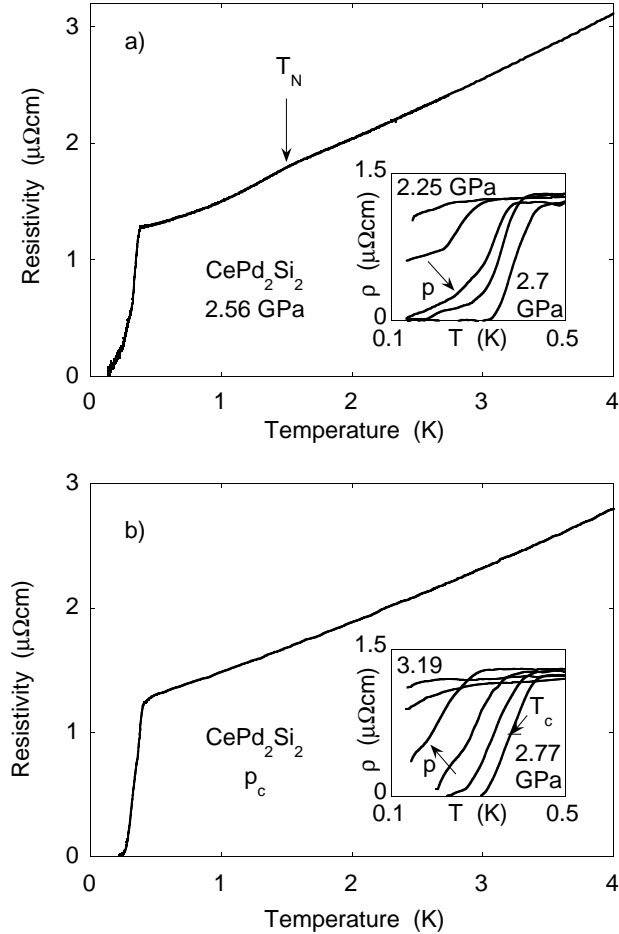,width=3.3in,clip=on}}
\par
\caption{Low temperature resistivity of CePd$_2$Si$_2$ measured along
the a-axis of a sample with $\mbox{RRR}=34$ (a) at 2.56
GPa, in the magnetic region of the phase diagram ($p<p_c$), where
$T_N$ is still visible by a kink in $\rho(T)$, and (b) at 2.8 GPa,
very close to $p_c$. The insets show the development of the
superconducting transition as the pressure is reduced (inset to (a)),
and increased (inset to (b)) away from $p_c$.}
\label{cpssupress}
\end{figure}

Thirdly, we can identify a small linear region in the resistivity just
above T$_N$ even at zero pressure. At higher pressures, this region
expands until, near the critical pressure, we observe a quasi-linear
resistivity over a wide range in temperature.  More precisely, the low
temperature form of the resistivity of CePd$_2$Si$_2$ follows a
power-law $\rho = \rho_0 + A T^x$, where x is a heuristic exponent
close to 1 (Fig.~\ref{cpsnormal}). This power-law dependence of the
resistivity characterises the sample response over nearly two orders
of magnitude in temperature up to about 40 K, where $\rho (T)$ crosses
over to a nearly constant high temperature value above 100 K
(Fig.~\ref{cpsnormal}). Comparing samples of different $\rho_0$ has
revealed a variation of the exponents in the range $1.1<x<1.5$ and a
general trend towards lower values for purer specimen, indicating a
possible limiting value of 1 for ideally pure samples
\cite{grosche98,grosche00}. This behaviour persists up to about 3.2
GPa, which is the pressure range accessible in this study, and no
return to a Fermi-liquid form of the resistivity could be observed at
low temperatures.

\begin{figure}[tb]
\centerline{\epsfig{figure=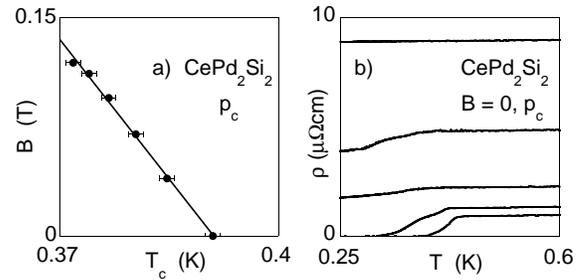,width=3.3in,clip=on}}
\par
\caption{(a) Magnetic-field dependence of $T_c$ for a sample of CePd$_2$Si$_2$,
with $\mbox{RRR}=34$, determined at the 80\%-point of the resistive
transition in small magnetic fields up to 0.12 T. The data indicate an
initial slope of $B'_{c2} \simeq -5 T/K$. (b) Low temperature
resistivity in CePd$_2$Si$_2$ samples of varying purity, indicated by
their low temperature limiting normal state resistivity, $\rho_0$. The
two purest samples, with RRR $= 55$ and 34 show full superconducting
transitions, while two more samples with RRR down to 10 show partial
transitions. No trace of superconductivity is seen in samples with $\mbox{RRR}
< 9$.}
\label{cpsfieldandsamples}
\end{figure}

The anomalous normal state characterised by the striking power-law
form of the resistivity close to $p_c$ is unstable towards
superconductivity at low temperatures (Fig.~\ref{cpssupress}).
Superconductivity in CePd$_2$Si$_2$ is inferred from the full loss of
resistivity in two samples, with residual resistance ratios (RRR) 34
and 55. Partial transitions were observed in two further samples with
RRR 21 and 9.4, while no superconductivity was observed in samples
with values of RRR less than 9. Figure \ref{cpsfieldandsamples}b shows
the transitions of these samples at $p \simeq p_c$. Increasing RRR
gives rise to sharper transitions and higher values for $T_c$,
indicating the importance of impurity scattering in breaking up the
superconducting pairs. A study of the upper critical field as a
function of temperature up to a field of 0.12 T on the sample with
$RRR=34$ leads to the conclusion that the initial slope $B'_{c2}\equiv d
B_{c2}/d T$ must be much higher in CePd$_2$Si$_2$ than in conventional
superconductors (Fig.~\ref{cpsfieldandsamples}a). The high slope,
which is estimated to be of the order of 5 T/K, together with the low
transition temperature imply a short coherence length, which is
consistent with a high effective mass of the quasi-particles forming
Cooper pairs. A treatment analogous to that used in
\cite{rauchschwalbe86} gives $\xi^2 \simeq c (T_c B'_{c2})^{-1}$, where
the constant $c \simeq 5 \cdot 10^4 T \mbox{\AA}^2$.  With $T_c \simeq
0.4 K$ and $B'_{c2} \simeq 5 T/K$, we estimate the coherence length as
$\xi \sim 150 \mbox{\AA}$.

A detailed study of the phase boundaries $T_c (p)$ and $T_N (p)$ was
carried out for one of these samples (RRR=34)
(Fig.~\ref{cpsphases}). We find that $T_c$, taken at the 50\%-point of
the resistive transition, goes through a single maximum at $p = 2.71
GPa$, just below the extrapolated critical pressure $p_c ~ 2.86 GPa$.
The curve $T_c (p)$ is approximately symmetric around its maximum and
has a width of $\Delta p \simeq 0.8 GPa$. Measurements on the other
fully and partially superconducting samples at selected pressures
confirm this behaviour.

Full superconducting transitions can be observed even in the ordered
state, where indications of both phase transitions are present
in $\rho(T)$  (Fig.~\ref{cpssupress}). However, the intriguing
question, whether both transitions occur  in the same region of the
sample, or whether the two phases are mutually exclusive, cannot be
answered straightforwardly from measurements of the resistivity alone.

\begin{figure}[bt]
\centerline{\epsfig{figure=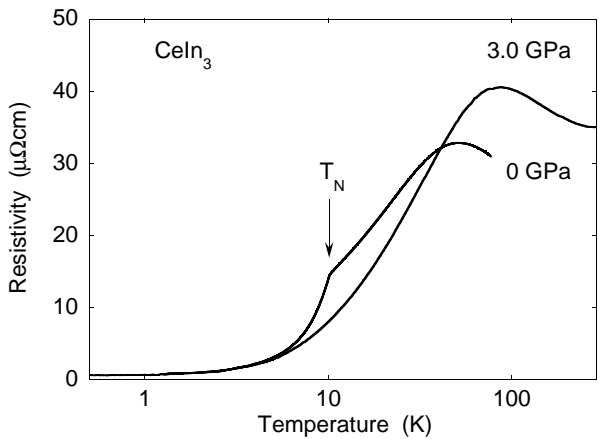,width=3.3in,clip=on}}
\par
\caption{
The resistivity of CeIn$_3$ at ambient pressure and at 3 GPa,
just above the critical pressure $p_c \simeq 2.6 GPa$, plotted vs. a
logarithmic temperature scale. The magnetic transition temperature,
$T_N$, marked by an arrow, is visible as a sudden change in slope and
is extracted, as in the case of CePd$_2$Si$_2$, from the midpoint of
$d\rho/dT$. The maximum of $\rho(T)$ is only weakly T-dependent,
compared to the case of CePd$_2$Si$_2$ and increases from about 70 K
to about 100 K over the pressure-range from 0 to 3 GPa.}
\label{cinresall}
\end{figure}

\subsection {CeIn$_3$}
CeIn$_3$ shows a qualitatively similar behaviour to CePd$_2$Si$_2$,
but a number of important differences are revealed on closer
inspection. Again, we identify a high temperature region, in which the
resistivity is nearly constant and goes through a weak maximum near 70
K (Fig.~\ref{cinresall}). With decreasing temperature, the resistivity
drops rapidly, and has a sharp anomaly at the N\'eel temperature of
about 10 K.  With increasing pressure, both temperatures shift, but
while $T_N$ is strongly pressure-dependent and can be suppressed at a
critical pressure $p_c \simeq 2.55 GPa$, the upper temperature scale
$T_{sf}$ appears to be much less pressure-dependent, when compared to
the case of CePd$_2$Si$_2$. Over the pressure range from 0 to 3 GPa,
the position of the maximum only changes from about 70 K to about 100
K.

In further contrast to CePd$_2$Si$_2$, the N\'eel temperature curve
$T_N (p)$ (Fig.~\ref{cinphases}) is seen to be concave down at higher
pressures, indicating a pressure dependence $T_N \propto (p_c-p)^x$ with
exponent $x < 1$ and more in line with theoretical predictions for the
role of 3D spin fluctuations in suppressing magnetic order.  It was
difficult to follow $T_N$ below 3 K, because the transition was less
clearly defined. It is likely that on approaching the critical
pressure the increasing slope of $T_N (p)$ leads to a wider
distribution of ordering temperatures within the sample due to
inhomogeneities of sample composition and pressure. This in turn would
give broader anomalies in the resistivity trace, making $T_N$ more
difficult to determine.

\begin{figure}[tb]
\centerline{\epsfig{figure=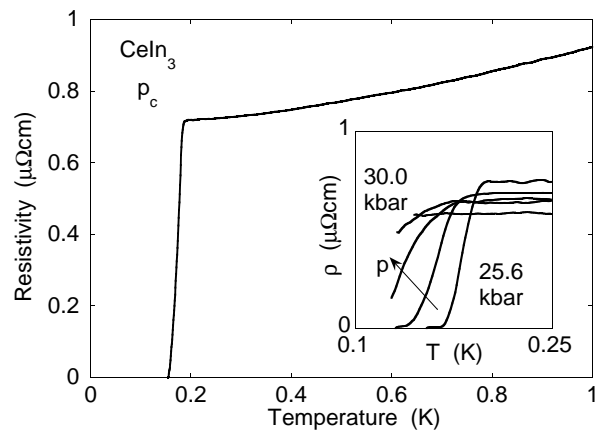,width=3.3in,clip=on}}
\par
\caption{Low temperature resistivity of CeIn$_3$ at $p_c \simeq 2.6
GPa$. The inset shows the development of the
superconducting transition as the pressure is 
increased away from $p_c$.}
\label{cinsupress}
\end{figure}

Close to the extrapolated critical pressure, sharp superconducting
transitions are seen below 200 mK (Fig.~\ref{cinsupress}).  The
superconducting region again reaches well into the ordered part of the
phase diagram and is confined within about 0.5 GPa, with a maximum at
2.55 GPa. 

%In contrast to CePd$_2$Si$_2$, the phase boundary $T_c (p)$
%of CeIn$_3$ is asymmetric around the maximum and $T_c$ decreases more
%quickly on the high pressure side than on the low pressure side.

The normal state resistivity at low temperatures approaches the form
predicted for a 3D antiferromagnet in the limit of very low
temperatures ($T \ll T_{sf}/RRR$), $\Delta\rho \propto T^{3/2}$ only
close to $p_c$ and at low temperatures (Fig.~\ref{cinnormal}).  The
power-law exponent x returns to the Fermi liquid value 2 more rapidly
than in CePd$_2$Si$_2$, as the system is tuned away from the critical
point, and does not lock into a constant value over a sizeable
temperature range.

\section {Discussion}

\subsection {Normal state resistivity}

\begin{figure}[tb]
\centerline{\epsfig{figure=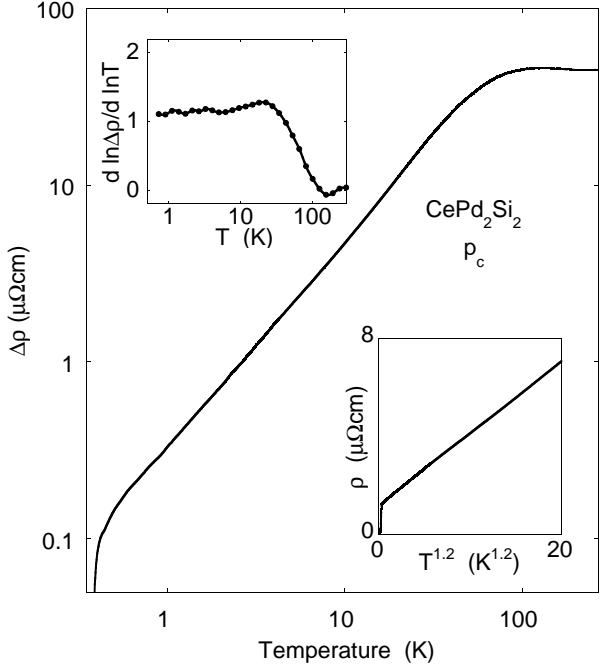,width=3.3in,clip=on}}
\par
\caption{The resistivity of CePd$_2$Si$_2$ near p$_c$. The main figure
shows $\Delta\rho\equiv\rho - \rho_0$ versus T on a doubly logarithmic
scale, where the residual resistivity $\rho_0\simeq 1.4\mu\Omega cm$
was estimated by a fitting procedure applied to the normal state, low
temperature data. The resistivity is essentially linear in $T^{1.2}$
over nearly two orders of magnitude in temperature down to the onset
of a superconducting transition near 0.4 K (lower inset) and crosses
over rapidly to a nearly constant value above about 40 K. The
temperature dependent exponent x(T), extracted by taking the
temperature logarithmic derivative of the resistivity, $x
\equiv\partial\log(\rho-\rho_0)/\partial \log T$ (upper inset),
exhibits a corresponding wide plateau close to 1.2 and a narrow
cross-over to 0 at high temperatures.}
\label{cpsnormal}
\end{figure}

The normal state data obtained in this study reveal
interesting differences between the tetragonal system CePd$_2$Si$_2$
and the cubic CeIn$_3$.
In CePd$_2$Si$_2$, the temperature dependence of the resistivity is
characterised over nearly two orders of magnitude in temperature by a
power-law $\rho=\rho_0+ A T^x$ with exponent x close to 1, and by a
rapid cross-over to a nearly constant resistivity at high temperatures
(Fig.~\ref{cpsnormal}).

\begin{figure}[tb]
\centerline{\epsfig{figure=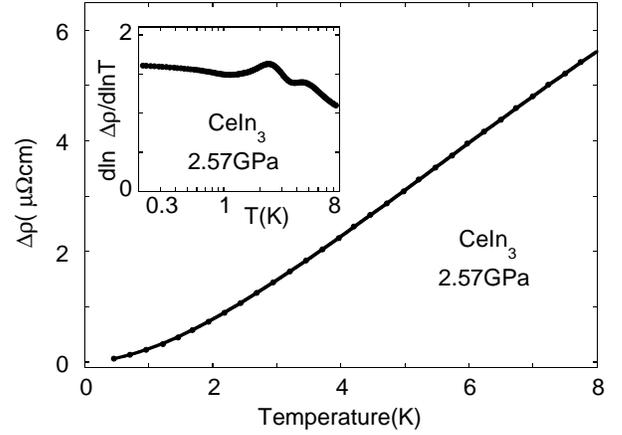,width=3.3in,clip=on}}
\par
\caption{The resistivity of CeIn$_3$ near p$_c \simeq 2.6 GPa$. The
main figure shows $\Delta\rho \equiv \rho - \rho_0$ versus T, where
the residual resistivity $\rho_0 \simeq 0.6 \mu\Omega cm$ was
estimated by a fitting procedure applied to the normal state, low
temperature data. The inset shows the temperature dependent exponent
$x(T)\equiv\partial\log(\rho-\rho_0)/\partial \log T$, which exhibits
a smooth cross-over towards a low-temperature limiting value close to
3/2.}
\label{cinnormal}
\end{figure}

These properties of the tetragonal metal CePd$_2$Si$_2$ and its
homologue CeNi$_2$Ge$_2$ contrast sharply with those of the cubic
antiferromagnet CeIn$_3$ \cite{walker97,mathur98}. In the latter, the
resistivity deviates from the Fermi liquid form only in a very narrow
pressure range near the critical pressure $p_c$. At $p_c$ and in low
magnetic fields the resistivity exponent, or more precisely $d \ln
(\rho-\rho_0) /d \ln T$, grows smoothly with decreasing temperature
and tends towards a value of about $3/2$ near 1 K
(Fig.~\ref{cinnormal}).
%By contrast, the observed exponents in CeNi$_2$Ge$_2$ and
%CePd$_2$Si$_2$ are closer to 1 and nearly constant over almost two
%orders of magnitude in temperature. 
For the origin of this intriguing difference, we may consider first
the known magnetic structure of CePd$_2$Si$_2$, which suggests a
frustrated spin coupling along the c-axis and hence a strongly
anisotropic spin fluctuation spectrum
\cite{grosche98,mathur98,grosche00} that may be more nearly 2D than 3D
as expected for CeIn$_3$. For a comparison with the predictions of the
standard magnetic interaction model we consider first the limit in
which regions about the `hot spots' on the Fermi surface (those states
whose self energies can exhibit a non-Fermi liquid form due to
antiferromagnetic spin fluctuations) dominate the temperature
dependence of the resistivity. This is expected to occur at very low
temperatures $T \ll T_{sf}/\mbox{RRR}$, when a short circuiting via
normal regions of the Fermi surface is blocked for example by
isotropic potential scattering via weak residual impurities. At a
continuous antiferromagnetic critical point one then expects in the
low temperature limit $\Delta\rho\sim T^{d/z}$, where $d$ is the
effective dimension of the spin fluctuation spectrum and $z$ is the
dynamical exponent normally taken to be 2 in our problem
\cite{lonzarich97}.  For this simplest model the resistivity exponent
is thus expected to be $3/2$ in 3D, in agreement with our observations
in cubic CeIn$_3$, and unity for 2D, in closer agreement to the
results in pure samples of the tetragonal CePd$_2$Si$_2$ and
CeNi$_2$Ge$_2$.  The precise degree of anisotropy can depend on
details of the microscopic near-neighbour exchange constants, giving
rise to the observed sample dependence of the power-law exponents.

Further evidence for a quasi two-dimensional character of the spin
fluctuation spectrum is drawn from the linear pressure-dependence of
$T_N$ observed in CePd$_2$Si$_2$ close to $p_c$.  In the magnetic
interaction model \cite{lonzarich97} this would indicate an effective
dimension close to 2, in agreement with the simplest interpretation of
the resistivity data and with recent neutron scattering work on the
related compound CeNi$_2$Ge$_2$ \cite{fak2000}.  
We note that by comparison $T_N(p)$ in CeIn$_3$ follows a power less
than 1 close to $p_c$, consistent with $d=3$ and with the observed low
temperature resistivity $\Delta\rho\sim T^{3/2}$. We stress that the
dimensionality $d$ refers to the magnetic fluctuation spectrum, not to
the carriers, which remain unconstrained to move in 3D. 

It may be instructive to relate the crystal and magnetic structures of
our antiferromagnetic {\em metals} to those of the extensively-studied
{\em insulators} K$_2$NiF$_4$ and KNiF$_3$ \cite{lines67}. Quasi-2D
magnetic behaviour has long been known to arise in the layered
perovskite K$_2$NiF$_4$, in which the magnetic moments of Ni occupy
bct positions and order with ${\bf Q}= [\frac{1}{2} \frac{1}{2} 0]$,
in exact analogy to the position and ordering of the Ce moments in
CePd$_2$Si$_2$. By contrast, its simple cubic counterpart KNiF$_3$
displays 3D magnetic behaviour and parallels CeIn$_3$ in the
arrangement of the magnetic moments and in the ordering wavevector
${\bf Q}=[\frac{1}{2} \frac{1}{2} \frac{1}{2}]$.

Because of competing contributions to the electrical conductivity from
hot and cold regions on the Fermi surface, the intermediate
temperature regime is more complex. The standard treatment outlined
above can be extended into this region \cite{hlubina95,rosch98}. Here,
one finds a temperature range in which exponents less than
$3/2$ can occur as a cross-over between high and low temperature forms
of $\rho(T)$ even for a 3D spin fluctuation spectrum, but it is too
early to tell whether or not it can account for all of the features we
observe, both in tetragonal CePd$_2$Si$_2$ and cubic CeIn$_3$, in a
consistent way. In particular, we note that the power-law resistivity
observed in CePd$_2$Si$_2$ at $p_c$ extends to nearly 40 K, i.e.
roughly $0.5$ T$_{sf}$, and lies far outside the predictive range of
the magnetic interaction model in its current form.

In a radical departure from the conventional model, an intuitive
description of this strange metallic state could be based on a more
extreme separation of the charge and spin degrees of freedom than is
present in current approaches \cite{coleman99,anderson97,si99b}.
Convincing evidence is emerging that in at least one other Ce-based
heavy fermion system, the CeCu$_{6-x}$Au$_x$ series, the quantum
critical point is accompanied by unexpected local spin dynamics
\cite{schroder98,schroder2000}.  Within the standard magnetic
interaction model the imaginary part of the wavevector and frequency
dependent susceptibility is of the form $\mbox{Im}\chi({\bf q},\omega)
\sim g({\bf q}) \omega^{\alpha}$ at sufficiently low $\omega (>0)$ and
$T$, where $g$ is some function of $\bf q$ and $\alpha = 1$. In the
standard model this value of $\alpha$ is expected to hold even in the
limit $T_N \rightarrow 0$ provided $d+z > 4$
\cite{lonzarich97}. However, in CeCu$_{6-x}$Au$_x$ it appears that in
the low-temperature limit the exponent $\alpha$ is anomalous and
significantly less than unity near the antiferromagnetic quantum
critical point.  Superficially, the slight deviation of $\alpha$ from
its theoretical value (together with the observed $\omega /
T$-scaling) might be seen as a minor correction, but on closer
examination, a sub-linear frequency dependence presents a major
challenge to our present understanding of metallic magnetism and could
imply a breakdown of Fermi liquid theory of an entirely different
order from that expected in any known model.

%Should this scenario carry over to other systems such as
%CePd$_2$Si$_2$, the physical picture would be that of spins
%fluctuating in a correlated electron bath, rather than that underlying
%the conventional scenarios, namely of a band magnet driven to a
%quantum critical point. In analogy to the electron-phonon problem, the
%fluctuating magnetic moments would then form an additional subsystem,
%giving rise, for example, to a separate contribution to the specific
%heat and not taking part in the formation of the Fermi surface.

Inelastic neutron scattering on the
pure, stoichiometric compound CeNi$_2$Ge$_2$ could decide whether such
behaviour can occur in the absence of disorder and whether it is,
ultimately, more widely spread in heavy fermion systems on the
threshold of magnetism.

%A further, more radical departure from a conventional description
%could focus on the local dynamics of the fluctuating spins. Neutron
%scattering results on the disordered system CeCu$_{6-x}Au_x$ indicate
%a possibly anomalous imaginary part to the magnetic response
%$Im(\chi_{\bf q \omega}) \sim \omega^\alpha, \alpha < 1$. 

%The influence of disorder is a further complicating factor in studies
%of the normal state properties, about which very little is known at
%this stage. Our findings indicate a clear trend towards smaller
%power-law exponents with increasing sample quality, as measured by
%$\rho_0$. The largest deviations from a Fermi-liquid type form
%$\Delta\rho \sim T^2$ are therefore seen in the samples with the
%smallest amounts of disorder, indicating that disorder is probably not
%the driving force for the 
%observed non-Fermi liquid behaviour in the
%resistivity.

\subsection{Superconductivity and magnetism}

The phase diagrams of CePd$_2$Si$_2$ and CeIn$_3$ show a striking
dependence of superconductivity on magnetism
(Figs.~\ref{cpsphases},\ref{cinphases}). The superconducting `bubble'
appears to be tacked onto the threshold of magnetic order, suggesting
that in these two systems, superconductivity is promoted, rather than
hindered, by the magnetic interaction which is at its largest near
$p_c$.  The discovery of a wide high-pressure superconducting phase
away from the critical point in the homologue to CePd$_2$Si$_2$,
CeNi$_2$Ge$_2$ \cite{grosche97}, indicates however, that the situation
in CePd$_2$Si$_2$ may be complicated by a further superconducting
state above $p_c$, beyond the range of our study.  For simplicity, our
discussion will focus on the narrow superconducting region near the
critical pressure, the existence of which has been independently
confirmed in both CePd$_2$Si$_2$ and CeIn$_3$
\cite{sheikin2000,demuer2000,knebel2000} and which appears most
closely associated to the presence of strong magnetic fluctuations.
Can we achieve at least a qualitative understanding of the salient
results emerging from these studies: the confinement of
superconductivity to a narrow region around $p_c$ and to samples with
purity levels exceeding a critical value?

%, and on the cubic material CeIn$_3$ in particular. 
%While
%more detailed calculations of $T_c$ for these materials are not
%completed, it may be helpful to discuss the influence of various
%material parameters on the formation of superconducting pairs.
%
%\begin{equation}
%T_c \sim \phi \Gamma exp(-\frac{1+\lambda+\lambda^*}{g\lambda})
%\label{Tc}
%\end{equation}
%In particular, we consider qualitatively the effect of four factors on the
%superconducting $T_c$:
%\begin{itemize}
%\item {The pair-breaking effect of impurities
%as soon as the  coherence length, $\xi$, outgrows the quasiparticle
%mean free path, $\ell$.}
%\item {The spin fluctuation
%bandwidth, loosely associated with $T_{sf}$ }
%\item {The strength $\lambda$ of the magnetic
%component of the quasiparticle interaction (defined through the
%mass-enhancement $\lambda \simeq \frac{m_\lambda}{m}$), relative to
%the strength of competing and pair-breaking interaction channels such
%as phonons or higher-lying branches of the spin fluctuation spectrum,
%($\lambda_1 \simeq \frac{m_{\lambda_1}}{m}$).}
%\item {A coupling constant $g$, which denotes the effectiveness of the
%magnetic interaction in forming Cooper pairs of a particular
%symmetry.}
%\end{itemize}

We consider first the role of impurity scattering in destroying
anisotropic superconductivity, as mean free paths shorter than
the superconducting coherence length, $\xi$, rapidly suppress $T_c$.
Because the sample
quality enters as the fraction of coherence length over mean free
path, longer coherence lengths limit superconductivity to purer
samples. Closeness to the quantum critical point, in turn, determines
the effective mass $m^*$, as evidenced by heat capacity measurements
on related systems, and therefore $\xi$.  The simplest mechanism
explaining the demise of superconductivity away from $p_c$ would
therefore lie in the increase of $\xi$ associated with the decrease in
$m^*$.  In the case of CePd$_2$Si$_2$, $\xi$ is estimated at $p_c$ to
be around 150 \AA, while the mean free path should go roughly as $\ell
\simeq 1500 \mbox{\AA} \mu\Omega cm/\rho_0$ \cite{rauchschwalbe86},
becoming comparable to $\xi$ for samples with a resistance ratio of
the order of 10. These numbers are consistent with our finding that full
transitions were confined to samples of CePd$_2$Si$_2$ with RRR larger
than 10.  Further, we can estimate the rate of decrease with pressure
of $\xi \propto 1/m^* \propto 1/\gamma$, where $\gamma$ is the
Sommerfeld coefficient of the specific heat, $C/T$, from pressure
measurements of $\gamma$ on the related material CeNi$_2$Ge$_2$. In
CeNi$_2$Ge$_2$, $\gamma$ drops at an initial rate of about $0.2
J/K^2mole/GPa$ from a zero-pressure value of about $0.4 J/K^2mole$
\cite{hellmann96}. Therefore, we can expect $\xi$ approximately to
double over a range of 1GPa, which should be sufficient to destroy
superconductivity in samples with $\mbox{RRR} \sim 20-30$.  These
numbers are in order-of-magnitude agreement with the observed width of
the superconducting region, but the fact that the cleanest sample,
with $\mbox{RRR} \simeq 60$, has a superconducting region no larger
than the $\mbox{RRR} \simeq 30$ sample suggests that further
mechanisms may be limiting superconductivity in CePd$_2$Si$_2$.  

One such influence may lie in the progressive competition from other
coupling mechanisms, as the relative strength of the magnetic
interaction decreases with distance from the critical point.  Within
the magnetic interaction model, superconductivity results from the
dominance of the magnetic channel over all other channels of the
quasiparticle interaction at the critical point. Moving away from
$p_c$, however, T$_c$ can diminish even in the absence of impurities,
as competing channels, which may favour different pairing symmetries,
become sufficiently strong to alter $T_c$ dramatically.

% $\lambda_1$ comparable to $\lambda$, and would
%rapidly suppress $T_c$.

While a similar mechanism may explain the confinement of
superconductivity in CeIn$_3$ close to $p_c$ and to pure samples,
there is an interesting discrepancy between the maximum values of
$T_c$ in the two materials. Despite higher levels of purity and a
similar value for $T_{sf}$, CeIn$_3$ has only half the $T_c$ of
CePd$_2$Si$_2$.  The size of $T_c$ depends critically on the coupling
parameter $\lambda_\Delta$, which involves the Cooper pair
wavefunction and the strength of the quasiparticle interaction. In the
case of phonon-mediated superconductivity, $\lambda_\Delta$ reduces to
the mass-enhancement $\lambda = (m^*-m)/m$, but it is a more complicated
property in anisotropic superconductors.
In accordance with recent numerical comparisons of the two- and
three-dimensional case \cite{monthoux99,monthoux00}, we speculate
that CePd$_2$Si$_2$ has a more favourable optimum value for
$\lambda_\Delta$ due to the more anisotropic and possibly nearly
two-dimensional nature of its spin fluctuation spectrum inferred from
its normal state properties.

One might naively expect that right at the critical pressure, nearly static
modes around the ordering wavevector $\bf Q$ should be pairbreaking
and reduce $T_c$ to a local dip. Recent numerical calculations \cite
{monthoux99,monthoux00}, however, do not predict such a dip at least in the
antiferromagnetic case, and indeed the data on both materials show no
evidence of a minimum at $p_c$. 

\subsection {Implications for CeCu$_2$Si$_2$ and other rare-earth-
compounds}

Magnetically mediated superconductivity may be expected to occur quite
generally in nearly magnetic metals of sufficiently high purity. For
many years, the scarcity of examples of superconductivity in 4f heavy
fermion systems that appeared to be on the border of magnetism and
that could be prepared in pure form seemed to contradict this
expectation. After extensive investigations over nearly 2 decades
superconductivity appeared to be limited solely to CeCu$_2$Si$_2$ and
its twin CeCu$_2$Ge$_2$. The new findings in CePd$_2$Si$_2$,
CeRh$_2$Si$_2$ \cite{movshovich96} and CeIn$_3$ may help to shed light
on the origin of this scarcity and to open the way to the discovery of
many new examples.

The key point is that superconductivity may indeed be ubiquitous in
heavy fermion systems, as anticipated by the magnetic interaction
model, but often only in very pure samples in a narrow range of
lattice density or pressure on the very edge of magnetic order. A
rapid collapse of superconductivity away from the critical density, or
critical pressure $p_c$, may appear surprising but it is not
incompatible with the magnetic interaction model for nearly
antiferromagnetic metals. Within this model, $T_c$ can fall with
increasing $|p-p_c|$ because of (i) the decrease in the ratio of the
mean free path to the superconducting coherence length as the
quasiparticle mass renormalistation factor falls, (ii) the decrease in
the strength and the non-locality of magnetic interactions and thus
the collapse of the pairing parameter $\lambda_\Delta$, and (iii) the
growing competition of other quasiparticle interaction channels.

Evidence for this precarious existence of superconductivity on the
border of antiferromagnetism around {\em ambient} pressure has been
seen in pure samples of CeNi$_2$Ge$_2$
\cite{grosche98,gegenwart99,braithwaite00}. In other nearly magnetic
systems, however, the same phenomena may only be observed via tuning
of the lattice density. This prescription has led to the discovery of
superconductivity not only in the above systems, but more recently
also in, CeCu$_2$ \cite{vargoz97}, CeCu$_5$Au \cite{wilhelm99},
CeRhIn$_5$ and CeIrIn$_5$ \cite{hegger2000}.

The crucial parameter in this approach, the `closeness' to the
critical point, needs to be defined more accurately to check whether
this scenario can apply even to the extreme case of {\em
CeCu$_2$Si$_2$}.  Amongst the many mechanisms under discussion, an
intriguingly simple possibility is that CeCu$_2$Si$_2$ is nearly
critical over a wide range in pressure, simply extending the
phenomenon observed in CeIn$_3$ and CePd$_2$Si$_2$ up to about 10 GPa,
in analogy to recent findings on CeNi$_2$Ge$_2$ and CePd$_2$Si$_2$
\cite{grosche98,grosche00,grosche97,lister96,agarwal00,braithwaite00}.  A
detailed analysis of the evolution of the normal state properties of
CeCu$_2$Si$_2$ with high hydrostatic pressure is under way to
see whether deviations from Fermi-liquid behaviour indeed extend over
the entire superconducting range.

However, even if the correlation length decreases with distance from
the quantum critical point, the mass-enhancement associated with the
magnetic interaction channel can still be large enough to allow
magnetically mediated superconductivity, if competing channels are
relatively weak and if the spin-fluctuation spectrum is soft over a
wide range in momentum-space. Low-lying modes over a large volume in
{\bf q}-space loosely imply a low dimensionality of the spin
fluctuation spectrum. They could arise from a nearly two-dimensional
character of the spin fluctuation spectrum, as is discussed for the
case of CePd$_2$Si$_2$, or an isotropic but weak dispersion of the
relaxation rate, $\Gamma_{\bf q}$. This dispersion can be estimated
from the spin-fluctuation temperature, which for CeCu$_2$Si$_2$ is much
lower than in most of the other superconducting Ce-compounds listed
above.

%While the phase diagrams are broadly consistent with the notion of a
%magnetically mediated superconducting state, it is not yet clear,
%whether the same mechanism could apply to CeCu$_2$Si$_2$.  In contrast
%to CePd$_2$Si$_2$ and CeIn$_3$, here we have a very large
%superconducting region, $T_c$ is anomalously high and it even has a
%maximum about 4 GPa away from $p_c$.

In view of other low-dimensional superconducting correlated electron
systems, the former route is of particular interest. Quantitative,
numerical comparisons of the superconducting transition temperature
for two- and three-dimensional nearly antiferromagnetic metals indicate a
significant enhancement of $T_c$ for lower dimensionality
\cite{monthoux99,monthoux00}. As a guiding principle, this approach
leads naturally to searching for magnetically mediated
superconductivity in 2-D materials and would seem
consistent with the recent discovery of superconductivity at
surprisingly high $T_c$ in two new heavy fermion compounds, CeIrIn$_5$
and CeRhIn$_5$, which form in a layered structure and appear to be
two-dimensional relatives of the cubic system
CeIn$_3$. \cite{hegger2000}.

\section{Conclusion}
CePd$_2$Si$_2$ and CeIn$_3$ present a unique opportunity to
investigate the role of lattice structure and magnetic anisotropy on
the superconducting and normal state properties on the  threshold of
magnetism.

At $p_c$, both materials exhibit an anomalous normal state extending
from an upper temperature scale $T_{sf}$ of the order of 100K down to
the onset of superconductivity below 430 mK and 175 mK,
respectively. The behaviour of the resistivity of CePd$_2$Si$_2$ is
particularly striking: it exhibits a nearly
linear temperature dependence over about two orders of magnitude in
temperature.

The superconducting state is characterised by its closeness to the
border of antiferromagnetic order. This link is underlined by the
small size of the superconducting region in both  materials,
indicating a likely magnetic origin for the attractive interaction
responsible for the formation of Cooper pairs. The wide separation
between $T_{sf}$ and $T_c$ could make these systems amenable to a
theoretical treatment of the superconducting transition.

Superconductivity in rare-earth based heavy fermion metals is still a
rare phenomenon. The general phase diagram observed in CePd$_2$Si$_2$
and CeIn$_3$ may provide the quantitative understanding necessary to
unravel other strongly correlated electron systems, including CeCu$_2$Si$_2$.

\acknowledgments We wish to thank P. Agarwal, S. Brown, F. Carter,
P. Coleman, J. Flouquet, K. Haselwimmer, D. Khmelnitskii, S. Lister,
A. Millis, P. Monthoux, C.  Pfleiderer, S. Saxena, G. Sparn, F. Steglich and
A. Tsvelik.  This research has been supported by the EPSRC of the UK,
the EU and the Newton Trust in Cambridge.

%\begin{thebibliography}{10}
%\bibitem [*]{byline} Present address: MPI-CPfS (Chemical Solid
%State Physics), Bayreuther Str. 40, D-01219 Dresden, Germany
%\end{thebibliography}

\bibliography {ref} 

\begin{thebibliography}{10}
\bibitem{resscales}
For CePd$_2$Si$_2$, the a-axis resistivity at ambient pressure was scaled to
  the room-temperature value given in \cite{raymond99}, $\rho(300K)=47\mu
  \Omega cm$, while for CeIn$_3$, we assumed $\rho(300K)=20 \mu \Omega cm$ for
  the ambient pressure resistivity, following \cite{morin88}.
\end{thebibliography}

\bibliographystyle {prsty}

\end {document}